
\magnification=\magstephalf

\newbox\SlashedBox
\def\slashed#1{\setbox\SlashedBox=\hbox{#1}
\hbox to 0pt{\hbox to 1\wd\SlashedBox{\hfil/\hfil}\hss}#1}
\def\hboxtosizeof#1#2{\setbox\SlashedBox=\hbox{#1}
\hbox to 1\wd\SlashedBox{#2}}

\def\mathslashed#1{\setbox\SlashedBox=\hbox{$#1$}
\hbox to 0pt{\hbox to 1\wd\SlashedBox{\hfil/\hfil}\hss}#1}

\def\ifsmall{\iffalse}  
\def\titlepagefont{}  

\def\DefineTeXgraphics{%
\special{ps::[global] /TeXgraphics { } def}}  

\def\today{\ifcase\month\or January\or February\or March\or April\or May
\or June\or July\or August\or September\or October\or November\or
December\fi\space\number\day, \number\year}
\def\eatPrefix19{}
\def\Year{\expandafter\eatPrefix\the\year}
\newcount\hours \newcount\minutes
\def\monthname{\ifcase\month\or
January\or February\or March\or April\or May\or June\or July\or
August\or September\or October\or November\or December\fi}
\def\shortmonthname{\ifcase\month\or
Jan\or Feb\or Mar\or Apr\or May\or Jun\or Jul\or
Aug\or Sep\or Oct\or Nov\or Dec\fi}

\def\TimeStamp{\hours\the\time\divide\hours by60%
\minutes -\the\time\divide\minutes by60\multiply\minutes by60%
\advance\minutes by\the\time%
${\rm \shortmonthname}\cdot\if\day<10{}0\fi\the\day\cdot\the\year%
\qquad\the\hours:\if\minutes<10{}0\fi\the\minutes$}




\def\Title#1{%
\vskip 1in{\titlefont\centerline{#1}}\vskip .5in}

\def\Date#1{\leftline{#1}\tenrm\supereject%
\global\hsize=\hsbody\global\hoffset=\hbodyoffset%
\footline={\hss\tenrm\folio\hss}}

\newif\ifdraftmode
\newif\ifleftlabels  

\def\nolabels{\def\wrlabeL##1{}\def\eqlabeL##1{}\def\reflabeL##1{}}
\def\writelabels{\def\wrlabeL##1{\leavevmode\vadjust{\rlap{\smash%
{\line{{\escapechar=` \hfill\rlap{\sevenrm\hskip.03in\string##1}}}}}}}%
\def\eqlabeL##1{{\escapechar-1\rlap{\sevenrm\hskip.05in\string##1}}}%
\def\reflabeL##1{\noexpand\rlap{\noexpand\sevenrm[\string##1]}}}
\def\writeleftlabels{\def\wrlabeL##1{\leavevmode\vadjust{\rlap{\smash%
{\line{{\escapechar=` \hfill\rlap{\sevenrm\hskip.03in\string##1}}}}}}}%
\def\eqlabeL##1{{\escapechar-1%
\rlap{\sixrm\hskip.05in\string##1}%
\llap{\sevenrm\string##1\hskip.03in\hbox to \hsize{}}}}%
\def\reflabeL##1{\noexpand\rlap{\noexpand\sevenrm[\string##1]}}}
\nolabels

\newdimen\fullhsize
\newdimen\hstitle
\hstitle=\hsize 
\newdimen\hsbody
\hsbody=\hsize 
\newdimen\hbodyoffset
\hbodyoffset=\hoffset 
\newbox\leftpage
\def\abstract#1{#1}
\def\rotated{\special{ps: landscape}
\magnification=1000  
\baselineskip=14pt
\global\hstitle=9truein\global\hsbody=4.75truein
\global\vsize=7truein\global\voffset=-.31truein
\global\hoffset=-0.54in\global\hbodyoffset=-.54truein
\global\fullhsize=10truein
\def\DefineTeXgraphics{%
\special{ps::[global]
/TeXgraphics {currentpoint translate 0.7 0.7 scale
              -80 0.72 mul -1000 0.72 mul translate} def}}
\let\lr=L
\def\ifsmall{\iftrue}
\def\titlepagefont{\twelvepoint}
\trueseventeenpoint
\def\almostshipout##1{\if L\lr \count1=1
      \global\setbox\leftpage=##1 \global\let\lr=R
   \else \count1=2
      \shipout\vbox{\hbox to\fullhsize{\box\leftpage\hfil##1}}
      \global\let\lr=L\fi}

\output={\ifnum\count0=1 
 \shipout\vbox{\hbox to \fullhsize{\hfill\pagebody\hfill}}\advancepageno
 \else
 \almostshipout{\leftline{\vbox{\pagebody\makefootline}}}\advancepageno
 \fi}

\def\abstract##1{{\leftskip=1.5in\rightskip=1.5in ##1\par}} }

\def\linemessage#1{\immediate\write16{#1}}

\global\newcount\secno \global\secno=0
\global\newcount\appno \global\appno=0
\global\newcount\meqno \global\meqno=1
\global\newcount\subsecno \global\subsecno=0
\global\newcount\figno \global\figno=0

\newif\ifAnyCounterChanged
\let\terminator=\relax
\def\normalize#1{\ifx#1\terminator\let\next=\relax\else%
\if#1i\aftergroup i\else\if#1v\aftergroup v\else\if#1x\aftergroup x%
\else\if#1l\aftergroup l\else\if#1c\aftergroup c\else%
\if#1m\aftergroup m\else%
\if#1I\aftergroup I\else\if#1V\aftergroup V\else\if#1X\aftergroup X%
\else\if#1L\aftergroup L\else\if#1C\aftergroup C\else%
\if#1M\aftergroup M\else\aftergroup#1\fi\fi\fi\fi\fi\fi\fi\fi\fi\fi\fi\fi%
\let\next=\normalize\fi%
\next}
\def\makeNormal#1#2{\def\doNormalDef{\edef#1}\begingroup%
\aftergroup\doNormalDef\aftergroup{\normalize#2\terminator\aftergroup}%
\endgroup}

\def\warnIfChanged#1#2{%
\ifundef#1
\else\begingroup%
\edef\oldDefinitionOfCounter{#1}\edef\newDefinitionOfCounter{#2}%
\ifx\oldDefinitionOfCounter\newDefinitionOfCounter%
\else%
\linemessage{Warning: definition of \noexpand#1 has changed.}%
\global\AnyCounterChangedtrue\fi\endgroup\fi}

\def\Section#1{\global\advance\secno by1\relax\global\meqno=1%
\global\subsecno=0%
\bigbreak\bigskip
\centerline{\twelvepoint \bf %
\the\secno. #1}%
\par\nobreak\medskip\nobreak}
\def\tagsection#1{%
\warnIfChanged#1{\the\secno}%
\xdef#1{\the\secno}%
\ifWritingAuxFile\immediate\write\auxfile{\noexpand\xdef\noexpand#1{#1}}\fi%
}
\def\section{\Section}
\def\Subsection#1{\global\advance\subsecno by1\relax\medskip %
\leftline{\bf\the\secno.\the\subsecno\ #1}%
\par\nobreak\smallskip\nobreak}
\def\tagsubsection#1{%
\warnIfChanged#1{\the\secno.\the\subsecno}%
\xdef#1{\the\secno.\the\subsecno}%
\ifWritingAuxFile\immediate\write\auxfile{\noexpand\xdef\noexpand#1{#1}}\fi%
}

\def\subsection{\Subsection}

\def\romappno{\uppercase\expandafter{\romannumeral\appno}}
\def\makeNormalizedRomappno{%
\expandafter\makeNormal\expandafter\normalizedromappno%
\expandafter{\romannumeral\appno}%
\edef\normalizedromappno{\uppercase{\normalizedromappno}}}
\def\Appendix#1{\global\advance\appno by1\relax\global\meqno=1\global\secno=0
\bigbreak\bigskip
\centerline{\twelvepoint \bf Appendix %
\romappno. #1}%
\par\nobreak\medskip\nobreak}
\def\tagappendix#1{\makeNormalizedRomappno%
\warnIfChanged#1{\normalizedromappno}%
\xdef#1{\normalizedromappno}%
\ifWritingAuxFile\immediate\write\auxfile{\noexpand\xdef\noexpand#1{#1}}\fi%
}
\def\appendix{\Appendix}

\def\eqn#1{\makeNormalizedRomappno%
\ifnum\secno>0%
  \warnIfChanged#1{\the\secno.\the\meqno}%
  \eqno(\the\secno.\the\meqno)\xdef#1{\the\secno.\the\meqno}%
     \global\advance\meqno by1
\else\ifnum\appno>0%
  \warnIfChanged#1{\normalizedromappno.\the\meqno}%
  \eqno({\rm\romappno}.\the\meqno)%
      \xdef#1{\normalizedromappno.\the\meqno}%
     \global\advance\meqno by1
\else%
  \warnIfChanged#1{\the\meqno}%
  \eqno(\the\meqno)\xdef#1{\the\meqno}%
     \global\advance\meqno by1
\fi\fi%
\eqlabeL#1%
\ifWritingAuxFile\immediate\write\auxfile{\noexpand\xdef\noexpand#1{#1}}\fi%
}
\def\defeqn#1{\makeNormalizedRomappno%
\ifnum\secno>0%
  \warnIfChanged#1{\the\secno.\the\meqno}%
  \xdef#1{\the\secno.\the\meqno}%
     \global\advance\meqno by1
\else\ifnum\appno>0%
  \warnIfChanged#1{\normalizedromappno.\the\meqno}%
  \xdef#1{\normalizedromappno.\the\meqno}%
     \global\advance\meqno by1
\else%
  \warnIfChanged#1{\the\meqno}%
  \xdef#1{\the\meqno}%
     \global\advance\meqno by1
\fi\fi%
\eqlabeL#1%
\ifWritingAuxFile\immediate\write\auxfile{\noexpand\xdef\noexpand#1{#1}}\fi%
}
\def\anoneqn{\makeNormalizedRomappno%
\ifnum\secno>0
  \eqno(\the\secno.\the\meqno)%
     \global\advance\meqno by1
\else\ifnum\appno>0
  \eqno({\rm\normalizedromappno}.\the\meqno)%
     \global\advance\meqno by1
\else
  \eqno(\the\meqno)%
     \global\advance\meqno by1
\fi\fi%
}
\def\mfig#1#2{\global\advance\figno by1%
\relax#1\the\figno%
\warnIfChanged#2{\the\figno}%
\edef#2{\the\figno}%
\reflabeL#2%
\ifWritingAuxFile\immediate\write\auxfile{\noexpand\xdef\noexpand#2{#2}}\fi%
}

\catcode`@=11 

\font\ninerm=cmr9
\font\eightrm=cmr8
\font\sixrm=cmr6

\def\loadtrueseventeenpoint{
 \font\seventeenrm=cmr10 at 17.28truept
 \font\seventeeni=cmmi10 at 17.28truept
 \font\seventeenbf=cmbx10 at 17.28truept
 \font\seventeenit=cmti10 at 17.28truept
 \font\seventeensl=cmsl10 at 17.28truept
 \font\seventeensy=cmsy10 at 17.28truept
}
\def\loadfourteenpoint{
\font\fourteenrm=cmr10 at 14.4pt
\font\fourteeni=cmmi10 at 14.4pt
\font\fourteenit=cmti10 at 14.4pt
\font\fourteensl=cmsl10 at 14.4pt
\font\fourteensy=cmsy10 at 14.4pt
\font\fourteenbf=cmbx10 at 14.4pt
}
\def\loadtruetwelvepoint{
\font\twelverm=cmr10 at 12truept
\font\twelvei=cmmi10 at 12truept
\font\twelveit=cmti10 at 12truept
\font\twelvesl=cmsl10 at 12truept
\font\twelvesy=cmsy10 at 12truept
\font\twelvebf=cmbx10 at 12truept
}

\font\ninei=cmmi9
\font\eighti=cmmi8
\font\sixi=cmmi6
\skewchar\ninei='177 \skewchar\eighti='177 \skewchar\sixi='177

\font\ninesy=cmsy9
\font\eightsy=cmsy8
\font\sixsy=cmsy6
\skewchar\ninesy='60 \skewchar\eightsy='60 \skewchar\sixsy='60

\font\ninebf=cmbx9
\font\eightbf=cmbx8
\font\sixbf=cmbx6

\font\ninett=cmtt9
\font\eighttt=cmtt8

\hyphenchar\tentt=-1 
\hyphenchar\ninett=-1
\hyphenchar\eighttt=-1

\font\ninesl=cmsl9
\font\eightsl=cmsl8

\font\nineit=cmti9
\font\eightit=cmti8


\newskip\ttglue
\def\tenpoint{\def\rm{\fam0\tenrm}%
  \textfont0=\tenrm \scriptfont0=\sevenrm \scriptscriptfont0=\fiverm
  \textfont1=\teni \scriptfont1=\seveni \scriptscriptfont1=\fivei
  \textfont2=\tensy \scriptfont2=\sevensy \scriptscriptfont2=\fivesy
  \textfont3=\tenex \scriptfont3=\tenex \scriptscriptfont3=\tenex
  \def\it{\fam\itfam\tenit}\textfont\itfam=\tenit
  \def\sl{\fam\slfam\tensl}\textfont\slfam=\tensl
  \def\bf{\fam\bffam\tenbf}\textfont\bffam=\tenbf \scriptfont\bffam=\sevenbf
  \scriptscriptfont\bffam=\fivebf
  \normalbaselineskip=12pt
  \let\sc=\eightrm
  \let\big=\tenbig
  \setbox\strutbox=\hbox{\vrule height8.5pt depth3.5pt width\z@}%
  \normalbaselines\rm}

\def\twelvepoint{\def\rm{\fam0\twelverm}%
  \textfont0=\twelverm \scriptfont0=\ninerm \scriptscriptfont0=\sevenrm
  \textfont1=\twelvei \scriptfont1=\ninei \scriptscriptfont1=\seveni
  \textfont2=\twelvesy \scriptfont2=\ninesy \scriptscriptfont2=\sevensy
  \textfont3=\tenex \scriptfont3=\tenex \scriptscriptfont3=\tenex
  \def\it{\fam\itfam\twelveit}\textfont\itfam=\twelveit
  \def\sl{\fam\slfam\twelvesl}\textfont\slfam=\twelvesl
  \def\bf{\fam\bffam\twelvebf}\textfont\bffam=\twelvebf
  \scriptfont\bffam=\ninebf
  \scriptscriptfont\bffam=\sevenbf
  \normalbaselineskip=12pt
  \let\sc=\eightrm
  \let\big=\tenbig
  \setbox\strutbox=\hbox{\vrule height8.5pt depth3.5pt width\z@}%
  \normalbaselines\rm}

\def\fourteenpoint{\def\rm{\fam0\fourteenrm}%
  \textfont0=\fourteenrm \scriptfont0=\tenrm \scriptscriptfont0=\sevenrm
  \textfont1=\fourteeni \scriptfont1=\teni \scriptscriptfont1=\seveni
  \textfont2=\fourteensy \scriptfont2=\tensy \scriptscriptfont2=\sevensy
  \textfont3=\tenex \scriptfont3=\tenex \scriptscriptfont3=\tenex
  \def\it{\fam\itfam\fourteenit}\textfont\itfam=\fourteenit
  \def\sl{\fam\slfam\fourteensl}\textfont\slfam=\fourteensl
  \def\bf{\fam\bffam\fourteenbf}\textfont\bffam=\fourteenbf%
  \scriptfont\bffam=\tenbf
  \scriptscriptfont\bffam=\sevenbf
  \normalbaselineskip=17pt
  \let\sc=\elevenrm
  \let\big=\tenbig
  \setbox\strutbox=\hbox{\vrule height8.5pt depth3.5pt width\z@}%
  \normalbaselines\rm}

\def\seventeenpoint{\def\rm{\fam0\seventeenrm}%
  \textfont0=\seventeenrm \scriptfont0=\fourteenrm \scriptscriptfont0=\tenrm
  \textfont1=\seventeeni \scriptfont1=\fourteeni \scriptscriptfont1=\teni
  \textfont2=\seventeensy \scriptfont2=\fourteensy \scriptscriptfont2=\tensy
  \textfont3=\tenex \scriptfont3=\tenex \scriptscriptfont3=\tenex
  \def\it{\fam\itfam\seventeenit}\textfont\itfam=\seventeenit
  \def\sl{\fam\slfam\seventeensl}\textfont\slfam=\seventeensl
  \def\bf{\fam\bffam\seventeenbf}\textfont\bffam=\seventeenbf%
  \scriptfont\bffam=\fourteenbf
  \scriptscriptfont\bffam=\twelvebf
  \normalbaselineskip=21pt
  \let\sc=\fourteenrm
  \let\big=\tenbig
  \setbox\strutbox=\hbox{\vrule height 12pt depth 6pt width\z@}%
  \normalbaselines\rm}

\def\ninepoint{\def\rm{\fam0\ninerm}%
  \textfont0=\ninerm \scriptfont0=\sixrm \scriptscriptfont0=\fiverm
  \textfont1=\ninei \scriptfont1=\sixi \scriptscriptfont1=\fivei
  \textfont2=\ninesy \scriptfont2=\sixsy \scriptscriptfont2=\fivesy
  \textfont3=\tenex \scriptfont3=\tenex \scriptscriptfont3=\tenex
  \def\it{\fam\itfam\nineit}\textfont\itfam=\nineit
  \def\sl{\fam\slfam\ninesl}\textfont\slfam=\ninesl
  \def\bf{\fam\bffam\ninebf}\textfont\bffam=\ninebf \scriptfont\bffam=\sixbf
  \scriptscriptfont\bffam=\fivebf
  \normalbaselineskip=11pt
  \let\sc=\sevenrm
  \let\big=\ninebig
  \setbox\strutbox=\hbox{\vrule height8pt depth3pt width\z@}%
  \normalbaselines\rm}

\def\eightpoint{\def\rm{\fam0\eightrm}%
  \textfont0=\eightrm \scriptfont0=\sixrm \scriptscriptfont0=\fiverm%
  \textfont1=\eighti \scriptfont1=\sixi \scriptscriptfont1=\fivei%
  \textfont2=\eightsy \scriptfont2=\sixsy \scriptscriptfont2=\fivesy%
  \textfont3=\tenex \scriptfont3=\tenex \scriptscriptfont3=\tenex%
  \def\it{\fam\itfam\eightit}\textfont\itfam=\eightit%
  \def\sl{\fam\slfam\eightsl}\textfont\slfam=\eightsl%
  \def\bf{\fam\bffam\eightbf}\textfont\bffam=\eightbf \scriptfont\bffam=\sixbf%
  \scriptscriptfont\bffam=\fivebf%
  \normalbaselineskip=9pt%
  \let\sc=\sixrm%
  \let\big=\eightbig%
  \setbox\strutbox=\hbox{\vrule height7pt depth2pt width\z@}%
  \normalbaselines\rm}

\def\tenbig#1{{\hbox{$\left#1\vbox to8.5pt{}\right.\n@space$}}}
\def\ninebig#1{{\hbox{$\textfont0=\tenrm\textfont2=\tensy
  \left#1\vbox to7.25pt{}\right.\n@space$}}}
\def\eightbig#1{{\hbox{$\textfont0=\ninerm\textfont2=\ninesy
  \left#1\vbox to6.5pt{}\right.\n@space$}}}

\def\footnote#1{\edef\@sf{\spacefactor\the\spacefactor}#1\@sf
      \insert\footins\bgroup\eightpoint
      \interlinepenalty100 \let\par=\endgraf
        \leftskip=\z@skip \rightskip=\z@skip
        \splittopskip=10pt plus 1pt minus 1pt \floatingpenalty=20000
        \smallskip\item{#1}\bgroup\strut\aftergroup\@foot\let\next}
\skip\footins=12pt plus 2pt minus 4pt 
\dimen\footins=30pc 

\newinsert\margin
\dimen\margin=\maxdimen
\def\titlefont{\seventeenpoint}
\loadtruetwelvepoint 
\loadtrueseventeenpoint
\catcode`\@=\active
\catcode`@=12  
\catcode`\"=\active

\def\eatOne#1{}
\def\ifundef#1{\expandafter\ifx%
\csname\expandafter\eatOne\string#1\endcsname\relax}
\def\notTrue{\iffalse}\def\isTrue{\iftrue}
\def\ifdef#1{{\ifundef#1%
\aftergroup\notTrue\else\aftergroup\isTrue\fi}}
\def\use#1{\ifundef#1\linemessage{Warning: \string#1 is undefined.}%
{\tt \string#1}\else#1\fi}


\global\newcount\refno \global\refno=1
\newwrite\rfile
\newlinechar=`\^^J
\def\ref#1#2{\the\refno\nref#1{#2}}
\def\nref#1#2{\xdef#1{\the\refno}%
\ifnum\refno=1\immediate\openout\rfile=refs.tmp\fi%
\immediate\write\rfile{\noexpand\item{[\noexpand#1]\ }#2.}%
\global\advance\refno by1}
\def\lref#1#2{\the\refno\xdef#1{\the\refno}%
\ifnum\refno=1\immediate\openout\rfile=refs.tmp\fi%
\immediate\write\rfile{\noexpand\item{[\noexpand#1]\ }#2\semi}%
\global\advance\refno by1}
\def\cref#1{\immediate\write\rfile{#1\semi}}

\def\semi{;\hfil\noexpand\break}

\def\inputAuxIfPresent#1{\immediate\openin1=#1
\ifeof1\message{No file \auxfileName; I'll create one.
}\else\closein1\relax\input\auxfileName\fi%
}
\def\NPB{Nucl.\ Phys.\ B}

\def\PL{Phys.\ Lett.\ }
\def\ZPC{Z.\ Phys.\ C}

\newif\ifWritingAuxFile
\newwrite\auxfile
\def\SetUpAuxFile{%
\xdef\auxfileName{\jobname.aux}%
\inputAuxIfPresent{\auxfileName}%
\WritingAuxFiletrue%
\immediate\openout\auxfile=\auxfileName}

\def\L{\left(}\def\R{\right)}

\def\LB{\left[}\def\RB{\right]}

\def\bye{\par\vfill\supereject%
\ifAnyCounterChanged\linemessage{
Some counters have changed.  Re-run tex to fix them up.}\fi%
\end}


\def\L{\left(}
\def\R{\right)}

\def\c{\mskip 1mu\cdot\mskip 1mu }

\def\eps{\epsilon}

\def\th#1#2{\maketh{\vartheta}{#1}{#2}}

\def\dl^#1_#2{\delta^{#1}{}_{#2}}

\def\Li{\mathop{\rm Li}\nolimits}

\def\Ord{{\cal O}}

\catcode`@=11  
\def\meqalign#1{\,\vcenter{\openup1\jot\m@th
   \ialign{\strut\hfil$\displaystyle{##}$ && $\displaystyle{{}##}$\hfil
             \crcr#1\crcr}}\,}
\catcode`@=12  


\baselineskip 15pt
\overfullrule 0.5pt


\def\c{\,\cdot\,}

\def\L{\left(}\def\R{\right)}

\def\spa#1.#2{\left\langle#1\,#2\right\rangle}
\def\spb#1.#2{\left[#1\,#2\right]}
\def\lor#1.#2{\left(#1\,#2\right)}
\def\sand#1.#2.#3{%
\left\langle\smash{#1}{\vphantom1}^{-}\right|{#2}%
\left|\smash{#3}{\vphantom1}^{-}\right\rangle}
\def\sandp#1.#2.#3{%
\left\langle\smash{#1}{\vphantom1}^{-}\right|{#2}%
\left|\smash{#3}{\vphantom1}^{+}\right\rangle}
\def\sandpp#1.#2.#3{%
\left\langle\smash{#1}{\vphantom1}^{+}\right|{#2}%
\left|\smash{#3}{\vphantom1}^{+}\right\rangle}
\catcode`@=11  
\def\meqalign#1{\,\vcenter{\openup1\jot\m@th
   \ialign{\strut\hfil$\displaystyle{##}$ && $\displaystyle{{}##}$\hfil
             \crcr#1\crcr}}\,}
\catcode`@=12  


\SetUpAuxFile
\loadfourteenpoint
\nopagenumbers\hsize=\hstitle\vskip1in
\overfullrule 0pt
\hfuzz 35 pt
\vbadness=10001
%
%

\def\hf{{\textstyle{1\over2}}}

\def\al#1{\alpha_{#1}}

\def\Li{\mathop{\rm Li}\nolimits}

\def\Det{\hat\Delta}
\def\e{\epsilon}
\def\del{\partial}
\def\hf{{\textstyle{1\over2}}}

\def\"#1{{\accent127 #1}}
\def\Basic{{\hat I}}

\def\Int{I}\def\Poly#1{\LB #1\RB}
\def\generalPoly{P(\{a_i\})}

\def\lr{\leftrightarrow}
\def\m#1{\hat m^2_{#1}}
\def\M#1{\hat M^2_{#1}}

\def\th#1{\hat t_{#1}}

\def\detprime{{\rm det}^\prime}
\def\rg{r_\Gamma}
\def\li{{\rm Li_2}}


\noindent
hep-ph/9212308 \hfill SLAC--PUB--6001 (T)
\rightline{CERN-TH.6756/92}
\rightline{UCLA/92/42}
\rightline{December, 1992}

\leftlabelstrue

\vskip -1.0 in
\Title{Dimensionally Regulated One-Loop Integrals}

\centerline{Zvi Bern${}^{\flat}$}
\baselineskip12truept
\centerline{\it Department of Physics}
\centerline{\it University of California, Los Angeles}
\centerline{\it Los Angeles, CA 90024}
\centerline{\tt bern@physics.ucla.edu}

\smallskip\smallskip

\baselineskip17truept
\centerline{Lance Dixon${}^{\star}$}
\baselineskip12truept
\centerline{\it Stanford Linear Accelerator Center}
\centerline{\it Stanford, CA 94309}
\centerline{\tt lance@slacvm.slac.stanford.edu}

\smallskip \centerline{and} \smallskip

\baselineskip17truept
\centerline{David A. Kosower}
\baselineskip12truept
\centerline{\it Theory Division}
\centerline{\it CERN}
\centerline{\it CH-1211 Geneva 23}
\centerline{\it Switzerland}
\vskip 5pt\centerline{and}\vskip 5pt
\centerline{\it Service de Physique Th\'eorique de Saclay${}^{\dagger}$}
\centerline{\it Centre d'Etudes de Saclay}
\centerline{\it F-91191 Gif-sur-Yvette cedex, France}
\centerline{\tt kosower@amoco.saclay.cea.fr}

\vskip 0.2in\baselineskip13truept

\centerline{\bf Abstract}

\ifdraftmode
\vskip 5pt
\centerline{{\bf Draft}\hskip 10pt\TimeStamp}
\vskip 5pt
\centerline{{\bf NOT for reproduction}}
\fi

{\narrower
We describe methods for evaluating one-loop integrals in
$4-2\e$ dimensions.  We give a recursion relation that expresses the
scalar $n$-point integral as a cyclicly symmetric combination of
$(n-1)$-point integrals.  The computation of such integrals thus
reduces to the calculation of box diagrams ($n=4$).  The tensor
integrals required in gauge theory may be obtained by differentiating
the scalar integral with respect to certain combinations of the
kinematic variables.  Such relations also lead to differential
equations for scalar integrals.  For box integrals with massless
internal lines these differential equations are easy to solve.  }

\baselineskip17pt

\centerline{\sl Submitted to Physics Letters B}

\vfill
\vskip 0.1in
\noindent\hrule width 3.6in\hfil\break
${}^{\flat}$Research supported by the Texas National Research Laboratory
Commission grant FCFY9202.\hfil\break
${}^{\star}$Research supported by the Department of
Energy under grant DE-AC03-76SF00515.\hfil\break
${}^{\dagger}$Laboratory of the {\it Direction des Sciences de la Mati\`ere\/}
of the {\it Commissariat \`a l'Energie Atomique\/} of France.\hfil\break
\Date{}

\line{}


Many processes of interest at current and future $e^+e^-$ and hadron
colliders involve large numbers of final state particles.
Radiative corrections to these
processes are needed for precise tests of the standard model.
It is therefore useful to have techniques for evaluating one-loop
integrals where the number of external legs is large.
As the loop integrals appearing in
radiative corrections are often infrared and/or ultraviolet
divergent, it is desirable to regulate them by
performing them in $4-2\e$ dimensions.

In this Letter, we derive a relation between
the $n$-point and $(n-1)$-point one-loop integrals,
which for $n>4$ allows the recursive determination of
the general $n$-point scalar integral in $D=4-2\e$,
as a linear combination of box integrals ($n=4$),
provided only that the external momenta are restricted to lie in
four dimensions, and neglecting ${\cal O}(\e)$ corrections.
The required box integrals can generally be evaluated
in closed form through ${\cal O}(1)$,
that is to say, with all poles in $\e$ manifest,
and with all functions of the kinematic invariants expressed in terms of
logarithms and dilogarithms.  (A compact expression for the
general infrared-finite box integral has recently been given by
Denner, Nierste, and Scharf~[\ref\NewFourPoint{%
A. Denner, U. Nierste, and R. Scharf, \NPB{367:637 (1991)}}];
the infrared-divergent box integrals with all internal lines massless
are collected in ref.~[\ref\OurPent{%
Z. Bern, L. Dixon and D. A. Kosower, preprint SLAC--PUB--5947, SPhT/92--048,
UCLA--92--43}].)
Therefore, the higher-point integrals can now be represented in the
same closed form.
In a separate paper~[\OurPent],
we apply these techniques to determine explicitly the pentagon integral
with all external lines massless, or with one external mass.

Various authors~[%
\ref\Melrose{D. B. Melrose, Il Nuovo Cimento 40A:181 (1965)},%
\ref\HVintegrals{G. 't Hooft and M. Veltman, \NPB{153:365 (1979)}},%
\ref\vNV{W. van Neerven and J. A. M. Vermaseren,
\PL{137B:241 (1984)}},\ref\EarlyPentagon{
G. J. van Oldenborgh and J. A. M. Vermaseren, \ZPC{46:425 (1990)}\semi
G. J. van Oldenborgh, PhD thesis, University of Amsterdam (1990)\semi
A. Aeppli, PhD thesis, University of Z\"urich (1992)}]
have discussed the computation
of pentagon and higher-point integrals that can be evaluated in $D=4$
(i.e. that are infrared finite).
In particular, Melrose~[\Melrose] and independently
van~Neerven and Vermaseren~[\vNV]
have expressed the $D=4$ pentagon integral as a linear combination of
five $D=4$ box integrals, and the relation we find for $n=5$ may be
thought of as the dimensionally-regulated version of their equations.
References~[\Melrose,\vNV] also express the four-dimensional
$n$-point scalar integral for $n\geq6$
(with external momenta restricted to $D=4$)
as a sum of six $(n-1)$-point integrals; the derivation in ref.~[\vNV]
extends straightforwardly
to ($4-2\e$)-dimensional loop-momenta as well.
For $n>6$ these relations are of a somewhat different type than
the relations that we find.
We have been informed that Ellis,
Giele, and Yehudai~[\ref\EandG{R. K. Ellis,  W. T. Giele, and E. Yehudai,
to appear}]
have recently evaluated the $D=4-2\e$ pentagon integrals by an
independent technique.

In gauge theories, tensor integrals appear in which the $n$-point
integral may contain up to
$n$ powers of the loop momentum in the numerator of the integrand.
It is possible to perform a
Brown-Feynman~[\ref\BrownFeynman{L. M. Brown and R. P. Feynman,
Phys.\ Rev.\ 85:231 (1952)}] or
Passarino-Veltman~[\ref\Passarino{G. Passarino and M. Veltman,
 \NPB{160:151 (1979)}}] reduction of the integrand,
solving a system of algebraic equations to reduce the tensor integrals
to a linear combination of scalar integrals~[\ref\Stuart{R. G. Stuart,
Comp.\ Phys.\ Comm.\ 48:367 (1988)\semi
R. G. Stuart and A. Gongora, Comp.\ Phys.\ Comm.\ 56:337 (1990)}].
The framework developed here provides another method for computing
tensor integrals.  Feynman parametrization converts tensor
integrals into integrals where polynomials
in the Feynman parameters (of order $\leq n$)
are inserted into the numerator of the integrand.
(In the string-motivated technique~[\ref\SMT{
  Z. Bern and D. A. Kosower, Phys.\ Rev.\ Lett.\ 66:1669 (1991)\semi
  Z. Bern and D. A. Kosower, Nucl.\ Phys.\ B379:451, 1992\semi
  Z. Bern and D. A. Kosower, in {\it Proceedings of the PASCOS-91
  Symposium}, eds.\ P. Nath and S. Reucroft\semi
  Z. Bern, D. C. Dunbar, Nucl.\ Phys.\ B379:562, 1992}]
for evaluating QCD amplitudes such a representation is obtained
directly.)
We will see that such integrals may be obtained by differentiating
the scalar integral with respect to particular combinations of
the kinematic variables.

The basic integral we wish to evaluate is the dimensionally-regulated
one-loop scalar
integral with $n$ external momenta $k_i$,
$n$ external masses $m_i$ ($k_i^2=m_i^2$), and $n$ internal masses
$M_i$, $i=1,2,\ldots,n$:
$$
  {\cal I}_n\ =\ \mu^{2\e}
  \int {d^{4 -2\eps} p \over (2 \pi)^{4-2\eps}}
  { 1 \over \bigl( p^2-M_1^2 \bigr) \bigl( (p-p_1)^2 -M_2^2 \bigr)
    \cdots \bigl( (p-p_{n-1})^2 - M_n^2 \bigr) } \; ,
\eqn\momspacedefn
$$
where we take
$$
p_i^\mu\ \equiv\ \sum_{j=1}^i k_j^\mu,\qquad\qquad
p_0^\mu\ =\ p_n^\mu\ =\ 0.
\eqn\pidefn
$$

Introducing Feynman parameters in~(\use\momspacedefn),
${\cal I}_n$ becomes
$$
  {\cal I}_n\ =\ \mu^{2\e}\ (n-1)! \int_0^1  d^na_i\
  \int {d^{4 -2\eps} p \over (2 \pi)^{4-2\eps}}
  \ \delta (1-{\textstyle \sum_i} a_i)\
  \biggl[ \sum_{i=1}^n a_i \bigl( (p-p_{i-1})^2 - M_i^2 \bigr)
  \biggr]^{-n} \ ,
\eqn\intermediateint
$$
Completing the square in the denominator, Wick rotating,
and integrating out the loop momentum yields
$$
  I_n[1]\ =\ \Gamma(n-2+\eps) \int_0^1  d^na_i\
   \delta (1-{\textstyle \sum_i} a_i)\
  {1 \over [{\cal D}(a_i)]^{n-2+\eps} } \ ,
\eqn\QIntegral
$$
where the scalar denominator ${\cal D}(a_i)$ is
$$
  {\cal D}(a_i)\ =\ \biggl[ \sum_{i=1}^n a_i p_{i-1} \biggr]^2
   \ -\ \sum_{i=1}^n a_i (p_{i-1}^2 - M_i^2)\ ,
\eqn\Deqnone
$$
and we have defined
$$
  I_n[1]\ \equiv\ i\ (-1)^{n+1}\ (4 \pi)^{2-\eps}\ \mu^{-2\e}
  \ {\cal I}_n
\anoneqn$$
as a convenient normalization for the Feynman parametrized scalar
integral.  (The expression inside the brackets indicates the
Feynman-parameter polynomial in the numerator of the
integrand; non-trivial polynomials
correspond to tensor integrals.)
We rewrite the second set of terms in ${\cal D}(a_i)$
using $\sum_i a_i = 1$, in order
to make it homogeneous of degree two in the $a_i$
and symmetric in $a_i \lr a_j$:
$$
  {\cal D}(a_i)\ =\ \sum_{i,j=1}^n  S_{ij}\ a_i a_j\ ,
\eqn\Deqntwo
$$
where the matrix $S_{ij}$ is given by
$$
  S_{ij}\ =\ \hf ( M_i^2 + M_j^2 - p_{ij}^2 ).
\eqn\Sdefn
$$
Here $p_{ij}$ is the sum of $|i-j|$ adjacent momenta,
$$
  p_{ii}\ \equiv\ 0,\qquad\qquad
  p_{ij}\ \equiv\ p_{j-1}-p_{i-1}
  \ =\ k_i+k_{i+1}+\cdots+k_{j-1}\quad{\rm for}\ i<j.
\eqn\pijdefn
$$

Now we are in a position to derive some useful general relations.
We start by evaluating the following integral two different ways:
$$
  J_1\ \equiv\ i\ (-1)^{n+1}\ (4 \pi)^{2-\eps}
   \int {d^{4 - 2\eps} p \over (2 \pi)^{4-2\eps}}
     {p^2 - M_1^2 \over
   \bigl( p^2-M_1^2 \bigr) \bigl( (p-p_1)^2 -M_2^2 \bigr)
    \cdots \bigl( (p-p_{n-1})^2 - M_n^2 \bigr) }
    \; .
\eqn\twoways
$$
First, we cancel the numerator against the denominator,
yielding the $(n-1)$-point integral
$$
  J_1\ =\ - I_{n-1}^{(1)}[1] \; .
\eqn\NMinusOneResult
$$
In general, the notation $I_{n-1}^{(i)}[1]$
refers to an $(n-1)$-point integral
whose kinematics is `inherited' from its `parent' $I_n[1]$
by removing the propagator between legs $i-1$ and $i$.
Second, we evaluate $J_1$ as an $n$-point integral.
Feynman parametrizing the integral
and completing the square in the denominator in the usual way yields
$$
\eqalign{
  J_1\ &=\ i\ (-1)^{n+1}\ (4 \pi)^{2-\eps} (n-1)! \int_0^1
  d^na_i\ \delta (1-{\textstyle \sum_i} a_i)
   \int {d^{4 -2\eps} q \over (2 \pi)^{4-2\eps}} \cr
  &\qquad \times
   {q^2 + {\cal D} + \sum_{i=1}^n a_i ( p_{i-1}^2 - M_1^2 - M_i^2 )
   \over [q^2 - {\cal D}]^n } \; . \cr}
\eqn\QIntegralnew
$$
Next we integrate out the loop momentum in eq.~(\use\QIntegralnew)
(after a Wick rotation), and equate the result
to the $(n-1)$-point result~(\use\NMinusOneResult),
to get an equation for $I_n[a_i]$:
$$
\eqalign{
  -I_{n-1}^{(1)}[1]\ &=\ \bigl[ -\hf(4-2\e) + (n-3+\e) \bigr]
  \, I_n^{D=6- 2 \eps}[1]
  \ +\ \sum_{i=1}^n ( p_{i-1}^2 - M_1^2 - M_i^2 ) I_n[a_i] \cr
    &=\ (n-5+2\eps)\, I_n^{D=6- 2 \eps}[1]
    \ -\ 2 \sum_{i=1}^n S_{1i}\ I_n[a_i]\ .\cr}
\eqn\OneParamEqn
$$
In this equation we have rewritten the terms coming from
$q^2 + {\cal D}$ in the numerator of equation~(\use\QIntegralnew),
using the Feynman-parameter representation of the
scalar $n$-point integral in $D=6-2\e$ dimensions,
$$
  I_n^{D=6-2\e}[1]\ =\ \Gamma(n-3+\eps) \int_0^1  d^na_i\
   \delta (1-{\textstyle \sum_i} a_i)\
  {1 \over [{\cal D}(a_i)]^{n-3+\eps} } \ .
\eqn\QIntegralsix
$$
(This representation can be obtained from equation~(\use\QIntegral)
simply by letting $\e\to\e-1$, which shifts $D=4-2\e$ to $D=6-2\e$.)
The properties of $I_n^{D=6-2\e}[1]$ as $\e\to0$ will play
a role below.

Similarly, by considering the integral $J_i$ with $(p-p_{i-1})^2-M_i^2$
in the numerator, we find the set of equations
$$
  2 \sum_{j=1}^n S_{ij}\ I_n[a_j]
  \ =\ I_{n-1}^{(i)}[1]\ +\ (n-5+2\eps)\, I_n^{D=6-2\eps}[1]\ .
\eqn\OneParamEqns
$$
Solving these equations for $I_n[a_i]$ we get
$$
  I_n[a_i]\ =\ {1\over2} \Biggl[
     \sum_{j=1}^n S^{-1}_{ij}\ I_{n-1}^{(j)}[1]
    \ +\ (n-5+2\eps)\, c_i\ I_n^{D=6-2\eps}[1] \Biggr]\ ,
\eqn\InamixedeqnS
$$
where
$$
  c_i\ =\ \sum_{j=1}^n S^{-1}_{ij}.
\eqn\cidef
$$
Now sum equation~(\use\InamixedeqnS) over $i$ and use $\sum_i a_i=1$
to get
$$
 I_n[1]\ =\ {1\over2} \Biggl[
     \sum_{i=1}^n c_i\ I_{n-1}^{(i)}[1]
    \ +\ (n-5+2\eps)\, c_0\ I_n^{D=6-2\eps}[1] \Biggr]\ ,
\eqn\allNeqnS
$$
where
$$
  c_0\ =\ \sum_{i=1}^n c_i\ =\ \sum_{i,j=1}^n S^{-1}_{ij}.
\eqn\czerodef
$$
Finally, use equation~(\use\allNeqnS) to
eliminate $I_5^{D=6-2\e}$ from equation~(\use\InamixedeqnS), and
thereby obtain an equation
for the one-parameter integrals $I_n[a_i]$ in terms of
$I_{n-1}[1]$ and $I_n[1]$:
$$
  I_n[a_i]\ =\ {1\over2} \sum_{j=1}^n c_{ij}\ I_{n-1}^{(j)}[1]
   \ +\ {c_i\over c_0}\ I_n[1]\ ,
\eqn\InaeqnS
$$
where
$$
  c_{ij}\ =\ S^{-1}_{ij} - {c_ic_j\over c_0}\ .
\eqn\cijdef
$$

The external momenta as well as the loop-momenta in
equations~(\use\InamixedeqnS), (\use\allNeqnS) and (\use\InaeqnS) are
still ($4-2\eps$)-dimensional.
In computing one-loop corrections to physical processes,
one is generally interested in restricting the external momenta
to $D=4$.
On the other hand, for $n>6$ the coefficients $c_0$ and $c_i$ given in
equations~(\use\czerodef) and (\use\cidef) appear to be singular for $D=4$
kinematics. (The rank of the $n\times n$ matrix $S_{ij}$
is $n-6$ in $D=4$~[\Melrose], so for $n>6$, $S$ is not invertible.)
In fact $c_0$ and $c_i$ are nonsingular in the limit of $D=4$
kinematics ($c_0$ actually vanishes for $n>5$).
In order to see this,
and in order to see how to apply these results to tensor integrals,
it is useful
to perform two changes of variables:  first a change of integration
variables in the integral~(\use\QIntegral),
then a change of kinematic variables.

Following 't~Hooft and Veltman~[\HVintegrals],
we make the change of integration variables in equation~(\use\QIntegral),
$$
\eqalign{
a_i\ &=\ {\al{i} u_i\over \sum^n_{j=1} \al{j} u_j}\ ,\qquad
       {\rm\ no\ sum\ on\ }i,\cr
a_n\ &=\ {\al{n} \L 1-\sum^{n-1}_{j=1} u_j\R\over \sum^n_{j=1} \al{j} u_j}
  \ .\cr
}
\eqn\ChangeOfVar
$$
Assuming that all $\al{i}$ are real and positive (physical regions may be
obtained by analytic continuation), the integral becomes
$$
\Int_n\Poly{1}\ =\ \Gamma(n-2+\e)\, \L \prod_{j=1}^n \al{j}\R\,
        \int_0^1 d^nu_i\; {\delta\L 1 - \sum u_i \R\,
          \L\sum^n_{j=1} \al{j} u_j\R^{n-4+2\eps}
         \over \LB \sum_{i,j=1}^n \rho_{ij} u_i u_j \RB^{n-2+\eps}}\ ,
\eqn\uintegral
$$
where
$$
  \rho_{ij}\ =\ S_{ij}\ \alpha_i\alpha_j\ =\
  \hf\bigl(  M_i^2 + M_j^2 - p_{ij}^2 \bigr)
  \alpha_i\alpha_j\ .
\eqn\rhodefn
$$

The form~(\use\uintegral) for the integral will be the most useful
if we also make a change of kinematic variables such that
the scalar denominator (i.e. the matrix $\rho$)
no longer depends on the $n$ variables $\alpha_i$.
First we describe a linearly independent set of the original kinematic
variables.  The $n(n+1)/2$ Mandelstam variables
$s_{ij}\ \equiv\ (k_i+k_j)^2$ are not
linearly independent, due to $n$ momentum conservation relations.
Instead one may use the square of the sum of $\ell$ adjacent momenta,
$p_{i,i+\ell}^2\ =\ (k_i + k_{i+1} + \ldots + k_{i+\ell-1})^2$.
For $n$ odd, $\ell$ runs from 1 to $(n-1)/2$.
For $n$ even, $\ell$ runs from 1 to $n/2$, but for $\ell=n/2$ there
are only $n/2$, rather than $n$, such variables.
These $n(n-1)/2$ linear combinations of Mandelstam variables
form a linearly independent, cyclicly symmetric set.
We should add to this set the $n$ internal masses $M_i^2$.
(Alternatively one may use the coefficients $S_{i,i+\ell}$
and $S_{ii}$ appearing in the scalar denominator ${\cal D}(a_i)$.)
For example, for the hexagon ($n=6$)
the independent variables are $M_i^2$, $m_i^2$,
$s_{i,i+1}\ \equiv\ (k_i+k_{i+1})^2$,
for $i=1,2,3,4,5,6$ (all indices are taken mod 6),
along with $t_{i,i+1,i+2}\ \equiv\ (k_i+k_{i+1}+k_{i+2})^2$, for $i=1,2,3$.

The following is an example of a change of kinematic variables that
eliminates all $\alpha_i$-dependence from the scalar denominator:
$$
   \{ M_i, m_i, s_{i,i+1}, t_{i,i+1,i+2},\ldots \} \quad\to\quad
   \{ \alpha_i;\M i,\m i,\th i,\ldots \}\ ,
\eqn\kinematicchange
$$
where the new variables $\{ \alpha_i;\M i,\m i,\th i,\ldots \}$
are defined by
$$
\eqalign{
  M_i^2 \ &=\ - {\M i \over \alpha_i^2 }\ ,  \cr
  m_i^2 \ &=\ - {\m i \over \alpha_i \alpha_{i+1} }
   - {\M i \over \alpha_i^2} - {\M {i+1} \over \alpha_{i+1}^2}\ , \cr
  s_{i, i+1}\ &=\ - {1 \over \alpha_i \alpha_{i+2} }
   - {\M i \over \alpha_i^2} - {\M {i+2} \over \alpha_{i+2}^2}\ , \cr
  t_{i, i+1,i+2}\ &=\ - {\th i \over \alpha_i \alpha_{i+3} }
   - {\M i \over \alpha_i^2} - {\M {i+3} \over \alpha_{i+3}^2}\ , \cr
   &\ldots \hskip 4 cm (i= 1,2,\ldots,n), \cr }
\eqn\standardchange
$$
with all indices taken mod $n$.
This is certainly not the only change of variables possible,
and in some cases it is convenient to make other choices.
(Indeed, for the special cases $n=8,12,16,\ldots$,
equations~(\use\standardchange) are not a legitimate change of variables.
This problem can be cured by modifying slightly the
substitution for just two of the $s_{i,i+1}$, say
$s_{12}\ =\ -{ \lambda_1 \over \alpha_1 \alpha_3 }
     - { \M 1 \over \alpha_1^2 } - { \M 3 \over \alpha_3^2 }$,
and
$s_{23}\ =\ - { \lambda_2 \over \alpha_2 \alpha_4 }
     - { \M 2 \over \alpha_2^2 } - { \M 4 \over \alpha_4^2 }$,
so that the set of kinematic variables is now
$\{ \alpha_i;\lambda_1,\lambda_2,\M i,\m i,\th i,\ldots \}$.)

With this change of kinematic variables,
tensor integrals may
be calculated simply by differentiating the scalar integral with respect to
the $\al{i}$.  Consider the Feynman-parametrized integral
with an arbitrary monomial of degree $m$ inserted in the numerator,
$$
  I_n[a_{i_1}a_{i_2}\ldots a_{i_m}]\ \equiv\
  \Gamma(n-2+\eps) \int_0^1  d^na_i\
  \delta (1-{\textstyle \sum_i} a_i)\
  { a_{i_1}a_{i_2}\ldots a_{i_m} \over [{\cal D}(a_i)]^{n-2+\eps} }
  \ .
\eqn\uintegralpoly
$$
Using equation~(\use\uintegral) it can be represented as
$$
  I_n[a_{i_1}a_{i_2}\ldots a_{i_m}]\ =\
  { \Gamma(n-3-m+2\e) \over \Gamma(n-3+2\e) }
  \L \prod_{j=1}^n \al{j}\R\, \al{i_1}\ldots \al{i_m}
  {\del\over\del\al{i_1}} \cdots {\del\over\del\al{i_m}}
  \L{ I_n[1] \over \prod_{j=1}^n \al{j} }\R\ .
\eqn\unreduceddiff
$$
If we define the {\it reduced integrals}
$$\eqalign{
\Basic_n\Poly{\hat\generalPoly}
  \ &\equiv\ \biggl(\prod_{j=1}^n \al{j}\biggr)^{-1}
   \Int_n\Poly{P(\{a_i/\al{i}\})}\ , \cr
  \Basic_n\ &\equiv\ \Basic_n[1]
  \ =\ \biggl(\prod_{j=1}^n \al{j}\biggr)^{-1} \Int_n[1]\ , \cr
}\eqn\ReducedIntegral
$$
then we may write
$$
  \Basic_n[a_{i_1}a_{i_2}\ldots a_{i_m}]\ =\
  { \Gamma(n-3-m+2\e) \over \Gamma(n-3+2\e) }\
   {\del^m\Basic_n \over \del\al{i_1} \cdots \del\al{i_m}}\ .
\eqn\reduceddiff
$$

We now proceed to find simple expressions for the coefficients
$c_0$, $c_i$ and $c_{ij}$ appearing in equations~(\use\allNeqnS),
(\use\InaeqnS), and (\use\InamixedeqnS), in terms of the $\alpha_i$
variables.  To do this it is first useful to
examine how the Gram determinant of the $(n-1)$-vector
system associated with the $n$-point process depends on $\alpha_i$.
The Gram determinant is defined by
$$
  \Delta_n\ \equiv\ \detprime(2k_i\c k_j),
\eqn\unreducednGram
$$
where the prime signifies that one of the $n$ vectors $k_i$ is to be
omitted before taking the determinant; due to momentum conservation,
$\sum k_i = 0$, any one of the vectors may be
omitted.\footnote{${}^\dagger$}{
  The notation for, and normalization of, the Gram determinant in
  equation~(\use\unreducednGram) differ from other conventions in
  the literature, e.g. references~[\Melrose,\ref\ByK{E. Byckling and K.
Kajantie, {\tenit Particle Kinematics} (Wiley) (1973)}].}

Next we introduce the {\it rescaled} Gram determinant,
$$
  \Det_n\ \equiv\
  \Bigl( \prod_{\ell=1}^n \alpha_\ell^2 \Bigr) \Delta_n,
\eqn\nGramdefn
$$
which will turn out to have several useful properties.
Rewrite $\Det_n$ in terms of the variables $\alpha_i$ and the
matrix $\rho$ defined in equation~(\use\rhodefn),
after omitting $k_n$ in the definition of $\Delta_n$:
$$
\eqalign{
  \Det_n
  \ &=\ \Bigl( \prod_{\ell=1}^n \alpha_\ell^2 \Bigr)
      \det_{i,j\neq n}(2 k_i \c k_j)
  \ =\ \Bigl( \prod_{\ell=1}^n \alpha_\ell^2 \Bigr)
       \det_{i,j\neq n}(2 p_i \c p_j) \cr
  \ &=\ 2^{n-1} \Bigl( \prod_{\ell=1}^n \alpha_\ell^2 \Bigr)
      \det_{i,j\neq 1}
      \biggl( { \rho_{ij} \over \alpha_i\alpha_j }
            - { \rho_{i1} \over \alpha_i\alpha_1 }
            - { \rho_{1j} \over \alpha_1\alpha_j }
            + { \rho_{11} \over \alpha_1^2 } \biggr) \cr
  \ &=\ 2^{n-1}\ \alpha_1^2\
      \det_{i,j\neq 1}
      \biggl( \rho_{ij} - \rho_{i1}{\alpha_j \over \alpha_1 }
        - \rho_{1j}{\alpha_i \over \alpha_1 }
        + \rho_{11}{\alpha_i\alpha_j \over \alpha_1^2 } \biggr)\ . \cr}
\eqn\nGramexpr
$$
By omitting other vectors $k_i$ in the definition of $\Delta_n$,
it is easy to obtain $n$ different expressions for the same
quantity $\Det_n$,
$$
  \Det_n
  \ =\ 2^{n-1}\ \alpha_\ell^2\
      \det_{i,j\neq \ell}
      \biggl( \rho_{ij} - \rho_{i \ell}{\alpha_j \over \alpha_\ell }
        - \rho_{\ell j}{\alpha_i \over \alpha_\ell }
        + \rho_{\ell\ell}{\alpha_i\alpha_j \over \alpha_\ell^2 } \biggr)
    \ , \qquad\qquad \ell=1,2,\ldots,n.
\eqn\nGramrhodefn
$$ From equation~(\use\nGramrhodefn), and the fact that the matrix $\rho$
is independent of the $\alpha_i$,
it is clear that $\Det_n$ is homogeneous of degree 2 in the $\alpha_i$,
and that no $\alpha_i$ appears with a negative power in $\Det_n$.
So we may write
$$
  \Det_n\ =\ \sum_{i,j=1}^n \eta_{ij}\alpha_i\alpha_j,
\eqn\etadefn
$$
where $\eta_{ij}$ is independent of the $\alpha_i$.

We can relate the two matrices $\eta$ and $\rho$.
Consider first the diagonal element $\eta_{\ell\ell}$.
The $\alpha_\ell^2$ term in $\Det_n$ comes from taking only
the $\rho_{ij}$ terms in~(\use\nGramrhodefn),
and is given by
$$
  \eta_{\ell\ell}\,\alpha_\ell^2\ =\
  2^{n-1}\det_{i,j\neq\ell}( \rho_{ij} )\ \alpha_\ell^2
  \ =\ 2^{n-1} \det\rho\ (\rho^{-1})_{\ell\ell}\ \alpha_\ell^2
\eqn\ellellcoeff
$$ (where the second determinant is over all indices).
Similarly, we pick off the $\alpha_\ell\alpha_m$ terms
($m\neq\ell$) in~(\use\nGramrhodefn), by using a $\rho_{i \ell}$ or
$\rho_{\ell j}$ term in place of $\rho_{i m}$ or $\rho_{m j}$,
to get $\eta_{\ell m}\ =\ 2^{n-1}\det\rho\ (\rho^{-1})_{\ell m}$.
Thus $\eta$ is proportional to the inverse of $\rho$,
$$
  \rho\ =\ N_n\ \eta^{-1},\qquad\qquad
  \eta\ =\ N_n\ \rho^{-1},
\eqn\Nnetarhoeqn
$$
where the proportionality constant is
$
  N_n\ \equiv\ 2^{n-1}\det\rho.
$

Now we rewrite the coefficients $c_0$, $c_i$ and $c_{ij}$
in terms of the $\alpha_i$ variables.
First of all, using equations~(\use\rhodefn) and (\use\Nnetarhoeqn)
the matrix $S^{-1}$ is given by
$$
  S^{-1}_{ij}\ =\ \alpha_i\, \rho^{-1}_{ij}\, \alpha_j
  \ =\ {\alpha_i\, \eta_{ij}\, \alpha_j \over N_n}\ .
\eqn\Sinv
$$
Summing this equation over $i$ and/or $j$, and using
equation~(\use\etadefn) for $\Det_n$, plus the definition
$$
  \gamma_i\ \equiv\ \sum_{j=1}^n \eta_{ij} \alpha_j
   \ =\ {1\over2}{\del\Det_n\over\del\alpha_i}
  \biggr\vert_{{\rm non}-\alpha_i\ {\rm variables\ fixed}}\ ,
\eqn\gammadefn
$$
we find that
$$
\eqalign{
     c_0\ &=\ {\Det_n \over N_n}\ , \hskip 0.2in
   c_{i}\ =\ {\alpha_i\gamma_i \over N_n}\ ,\hskip 0.2in
  c_{ij}\ =\ {\alpha_i\alpha_j \over N_n}\,
  \left(\eta_{ij}-{\gamma_i\gamma_j\over\Det_n}\right)\ .  }
\eqn\cijalpha
$$

In terms of the reduced integrals defined in
equation~(\use\ReducedIntegral) --- which have simple differentiation
properties --- equations~(\use\allNeqnS), (\use\InaeqnS), and
(\use\InamixedeqnS) also take on a simpler form.
It is convenient to make the change of kinematic variables for the
integrals $I_{n-1}^{(i)}[1]$, such that the variables $\alpha_i$
are identical to those for the parent integral $I_n[1]$.
Then we have
$$
  \Basic_n\ =\ {1\over 2N_n} \Biggl[ \sum_{i=1}^n \gamma_i \,
   \Basic_{n-1}^{(i)}\ +\ (n-5+2\e) \,\Det_n \,
   \Basic_n^{D=6-2\e} \Biggr]\ ,
\eqn\reducedallNeqn
$$
$$
  {1\over n-4+2\e}\ {\del\Basic_n\over\del\alpha_i}\ =\ \Basic_n[a_i]
  \ =\ {1\over 2 N_n} \sum_{j=1}^n
  \Bigl( \eta_{ij} - {\gamma_i\gamma_j\over\Det_n} \Bigr)
   \Basic_{n-1}^{(j)} \ +\ {\gamma_i\over\Det_n} \Basic_n\ .
\eqn\reducedInaeqn
$$
$$
  {1\over n-4+2\e}\ {\del\Basic_n\over\del\alpha_i}\ =\ \Basic_n[a_i]
  \ =\ {1\over 2 N_n} \Biggl[ \sum_{j=1}^n
   \eta_{ij}\ \Basic_{n-1}^{(j)}
   \ +\ (n-5+2\e)\,\gamma_i\ \Basic_n^{D=6-2\e} \Biggr]\ .
\eqn\reducedInamixedeqn
$$

The simple equations~(\use\reducedallNeqn), (\use\reducedInaeqn), and
(\use\reducedInamixedeqn) are in some sense the main results of this letter.
All the equations derived so far are valid
in an arbitrary spacetime dimension $D=4-2\e$; that is,
we have not yet assumed that $\e$ is small, nor that the external
momenta lie in $D=4$.  However, the equations
have the most utility in the context of dimensional regularization,
i.e. for $D=4-2\e$ with $\e$ tending to zero.
We now discuss how to use the equations to recursively generate
one-loop $n$-point integrals, up to ${\cal O}(\e)$ corrections.

At first sight equation~(\use\reducedallNeqn) does not appear very useful,
due to the presence of the integral $I_n^{D=6-2\e}[1]$
on the right-hand-side.
However, for $n=5$ the coefficient of this term is of order $\e$;
and the integral $I_5^{D=6-2\e}[1]$ is finite as $\e\to0$, because
the $D=6$ scalar pentagon integral
possesses neither ultraviolet divergences nor infrared divergences
(soft or collinear).
So to order $\e$ the $D=4-2\e$ scalar pentagon is
given by a sum of five scalar boxes,
$$
  I_5[1]\ =\ {1\over2}\sum_{i=1}^5
    \left( \sum_{j=1}^5 S^{-1}_{ij} \right)\ I_4^{(i)}[1]
    \ +\ {\cal O}(\e)
   \ =\ {1\over 2N_5}\sum_{i=1}^5 \alpha_i \gamma_i \,
   I_4^{(i)}[1]\ +\ {\cal O}(\e).
\eqn\Neqfiveeqn
$$
It is easy to check that
the coefficients of the box integrals in this equation are identical
to those in the corresponding $D=4$ relation in ref.~[\Melrose];
both sets of coefficients are expressed in terms of matrix elements
of $S_{ij} = \hf(M_i^2+M_j^2-p_{ij}^2)$.
In ref.~[\vNV] the box coefficients in the $D=4$ relation were expressed
using Levi-Civita symbols; in ref.~[\OurPent] we show that the
coefficients are nevertheless the same as those in
equation~(\use\Neqfiveeqn).

For $n\geq6$, the coefficient of $I_n^{D=6-2\e}[1]$ in
equation~(\use\allNeqnS) is not of order $\e$.
On the other hand, if all external momenta are chosen to lie
in $D=4$ dimensions, then we can set $c_0=0$.  This is because
the Gram determinant $\Delta_n$ vanishes in $D=4$,
due to the linear dependence of the $(n-1)$ vectors
$k_1,k_2,\ldots,k_{n-1}$~[\Melrose,\use\ByK,\ref\Gram{%
V. E.\ Asribekov, Sov.\ Phys.\ -- JETP 15:394 (1962)\semi
N. Byers and C. N. Yang, Rev.\ Mod.\ Phys.\ 36:595 (1964)}].
In next-to-leading-order calculations, it is always possible to
put this restriction on the external kinematics.
We then get
$$
  I_n[1]\ =\ {1\over 2N_n} \sum_{i=1}^n \alpha_i \gamma_i \,
   I_{n-1}^{(i)}[1]\ ,\qquad\qquad n\geq6,\quad
   D=4\ {\rm external\ kinematics}.
\eqn\Ngeqsixeqn
$$
For $n=6$ when all integrals appearing in~(\use\Ngeqsixeqn) are finite,
this equation is identical to the non dimensionally-regulated result
of ref.~[\Melrose], and it can be shown to be equivalent
to the corresponding result of ref.~[\vNV], expressed in terms of
Levi-Civita symbols; the latter derivation extends straightforwardly to
$(4-2\e)$-dimensional loop momenta.

For $n>6$, equation~(\use\Ngeqsixeqn) differs somewhat from
the results of references~[\Melrose,\vNV] in that it contains
$n$, rather than six, $(n-1)$-point integrals.
In the reductions~(\use\Neqfiveeqn)
and (\use\Ngeqsixeqn), the cyclic symmetry of $I_n[1]$ is kept manifest.
A more important difference arises if one wishes to
extract tensor integrals via the differentiation formula~(\reduceddiff).
Namely, one should not restrict to four-dimensional external kinematics
until after carrying out the differentiation, and so the cyclicly
symmetric representation~(\use\reducedallNeqn), with unrestricted
kinematics, should be used as the starting point.

As mentioned above, for $n>6$ the representations~(\use\czerodef) and
(\use\cidef) of the coefficients $c_0$
and $c_i$ in terms of the original kinematic variables
$S_{ij}$ are problematic when the external momenta are restricted
to $D=4$.  This is because the rank of the $n\times n$ matrix $S_{ij}$
is $n-6$ in $D=4$~[\Melrose], so for $n>6$ the inverse $S^{-1}_{ij}$
does not exist.  On the other hand, the $\alpha_i$-representations
of $c_0$ and $c_i$ in equation~(\use\cijalpha)
are non-singular and well-defined for $D=4$ kinematics.
In summary, the combination of equations~(\use\Neqfiveeqn)
and (\use\Ngeqsixeqn) recursively determines the general one-loop
$n$-point scalar integral in $D=4-2\e$ dimensions as
a linear combination of box integrals.

Equation~(\use\allNeqnS) also has significance for $n\leq4$,
even though the term containing $I_n^{D=6-2\e}[1]$ in
the equation may no longer be neglected.
For example, for $n=4$ the decomposition~(\use\allNeqnS)
of $I_4^{D=4-2\e}[1]$ has the virtue of putting all the $\e\to0$
divergences into the triangle integrals $I_{3}^{(i)}[1]$, since the
$D=6$ scalar box is infrared and ultraviolet finite.
This is important for practical calculations
because infrared-divergent triangle integrals are
generally simple to evaluate analytically;
the infrared-finite triangles and the
$D=6$ scalar box can be evaluated numerically if necessary.

Equation~(\use\reducedInaeqn) gives a set of partial differential
equations for $n$-point scalar integrals.
For box integrals ($n=4$), one may also use
equation~(\use\reducedInamixedeqn), plus the finiteness of the $D=6$
scalar box, to simplify the differential equations
through ${\cal O}(\e)$:
$$
  {\del\Basic_4\over\del\alpha_i}
  \ =\ {\e \over N_4} \Biggl[
  \sum_{j=1}^4 \eta_{ij}\ \Basic_3^{(j)}
  \ +\ (-1+2\e)\,\gamma_i\ \Basic_4^{D=6-2\e} \Biggr]
  \ =\ {\e\over N_4} \sum_{j=1}^n \eta_{ij}\ \Basic_3^{(j)}
   \ +\ {\cal O}(\e)\ .
\eqn\boxPDEs
$$
In ref.~[\OurPent] we solve these differential equations
for box integrals with all internal lines massless,
but with nonzero masses for 0, 1, 2, or 3 external lines.
(Some of these integrals have been computed previously by
other techniques.)
Together with the infrared-finite box integral with all
four external lines massive~[\NewFourPoint],
these are the complete set of box integrals needed to
recursively determine the $n$-point integrals for
next-to-leading-order calculations in QCD with massless quarks.

For example, the box integral with one external
mass is needed to obtain the all-massless pentagon via
equation~(\use\Neqfiveeqn).  The solution to the differential
equations for the reduced box integral is~[\OurPent]
$$
  \Basic_4^{(i)}\ =\ 2\rg\ \Biggl[
   {(\alpha_{i+2}\alpha_{i-2})^\e \over \e^2}
   + \li\left( 1 - {\alpha_{i+1}\over\alpha_{i+2}} \right)
   + \li\left( 1 - {\alpha_{i-1}\over\alpha_{i-2}} \right)
   - {\pi^2\over6} \Biggr]\ +\ {\cal O}(\e),
\eqn\onemassbox
$$
where $\rg = \Gamma(1+\e)\Gamma^2(1-\e)/\Gamma(1-2\e)$,
and we have written the integral in terms
of a set of $\alpha_i$ kinematic variables that are appropriate for
the all-massless pentagon; the $\alpha_i$ are defined by
$$
  s_{i,i+1}\ =\ -{1\over\alpha_i\alpha_{i+2}}\ , \qquad
  i=1,2,\ldots,5\quad ({\rm mod}\ 5),
\eqn\fivealphadefn
$$
where $s_{i,i+1}$ are the five independent momentum invariants for the
pentagon.  The coefficients $\gamma_i$ are then
$$
 \gamma_i\ =\ \alpha_{i-2} - \alpha_{i-1} + \alpha_{i}
             - \alpha_{i+1} + \alpha_{i+2},
\eqn\fivegammadefn
$$
and the normalization constant is $N_5=1$.
Plugging equation~(\use\onemassbox) into the formula~(\use\Neqfiveeqn)
for the massless scalar pentagon, and using the dilogarithm identity
$\li(1-x)+\li(1-x^{-1}) = -\hf \ln^2 x$ to simplify the expression,
we find
$$
  \Basic_5^{\rm massless}\ =\ \rg\ \sum_{i=1}^5
 \alpha_i^{1+2\e} \biggl[ {1\over\e^2}
   + 2\ \li\left( 1 - {\alpha_{i+1}\over\alpha_i} \right)
   + 2\ \li\left( 1 - {\alpha_{i-1}\over\alpha_i} \right)
   - {\pi^2\over6} \biggr]\ +\ {\cal O}(\e).
\eqn\reducedpent
$$
This solution can also be obtained by solving the differential
equations~(\use\reducedInaeqn) or (\use\reducedInamixedeqn).
In terms of momentum invariants, the unreduced massless scalar
pentagon integral is
$$
\eqalign{
  I_5^{\rm massless}[1]\ &=\
  {\rg\ (-s_{12})^\eps (-s_{51})^\eps \over
 (-s_{23})^{1+\eps} (-s_{34} )^{1+\eps} (-s_{45})^{1+\eps} }
 \LB {1\over \eps^2} + 2\Li_2\Bigl(1- {s_{23} \over s_{51} }\Bigr)
+2\Li_2\Bigl( 1- {s_{45}\over s_{12}} \Bigr) -
{\pi^2 \over 6} \RB \cr
 &\qquad\ +\ \hbox{cyclic permutations}\ +\ \Ord(\eps). \cr}
\anoneqn
$$
$I_5^{\rm massless}[1]$ is manifestly real in the region where
all $s_{ij} < 0$; its value in physical regions can be obtained
by the usual prescription $s_{ij} \to s_{ij} + i\varepsilon$.

There are a few subtleties to obtaining tensor integrals via the
derivative formula~(\use\reduceddiff) when $n\geq4$.
Some of the subtleties are associated with the
$1/\e$ pole in $\Gamma(n-3-m+2\e)$ for $m\geq n-3$, where
$m$ is the degree of the Feynman parameter monomial being integrated.
It might appear that the calculation of such integrals to
${\cal O}(1)$ would require knowledge of the scalar integral
to ${\cal O}(\e)$; however, this is not the case.
These issues are treated in more detail in ref.~[\OurPent]; here
we simply note that equation~(\use\InaeqnS) for $n=4$ allows
the determination of $I_4[a_i]$ to ${\cal O}(1)$, given
$I_3[1]$ and $I_4[1]$ to ${\cal O}(1)$.
The integrals of monomials with degree greater than
1 can then be obtained to ${\cal O}(1)$ by differentiating the
integrals $I_4[a_i]$.  Similarly, for $n=5$ it is possible to
derive an equation that yields $I_5[a_ia_j]$ to ${\cal O}(1)$,
given $I_3[1]$ and $I_4[1]$ (or equivalently $I_4^{D=6}[1]$)
to ${\cal O}(1)$.
In this case, the $D=6$ scalar pentagon $I_5^{D=6}[1]$
appears on the right-hand-side of the equation
as well, but it is possible to
show that for Feynman parameter polynomials that arise from
tensor integrals in the loop-momentum, the coefficient of
$I_5^{D=6}$ always vanishes.
The integrals of monomials with degree greater than 2
can be obtained by differentiating $I_5[a_ia_j]$.  One also
uses equation~(\use\reducedInaeqn) for $n=5$ and $\e\to\e-1$ to
eliminate the derivatives $\del\Basic_5^{D=6-2\e}/\del\alpha_i$
in favor of box integrals plus the scalar integral
$\Basic_5^{D=6-2\e}$.  After doing this,
the coefficient of the $D=6$ scalar pentagon integral will always
vanish, for Feynman parameter polynomials
arising from tensor integrals in the loop-momentum.

For tensor integrals with $n\geq6$
there is another subtlety, associated with the appearance of
$\Det_n$ in the denominator of the coefficient $c_{ij}$ in
equation~(\use\cijalpha), since $\Det_n=0$ for $D=4$ kinematics and
$n\geq6$.  On the other hand, the scalar integrals are manifestly
free of singularities as $\Det_n\to0$ (so long as
$s_{i,i+1} \not\rightarrow 0$), and by considering
the tensor integrals in terms of loop-momenta, one can see that
they also can have no singularity as $\Det_n\to0$.
Therefore the $1/\Det_n$ factors that appear in some representations
of the tensor integrals (such as equation~(\use\reducedInaeqn)),
should cancel out when all quantities are
evaluated explicitly.  We have checked that this is true for insertions
of up to two Feynman parameters.

In conclusion, we have derived simple equations relating $I_n[1]$,
the one-loop $n$-point scalar integral in $D=4-2\e$, to
$I_{n-1}^{(i)}[1]$ and the ($6-2\e$)-dimensional integral
$I_n^{D=6-2\e}[1]$.  In the context of dimensional regulation
these equations may be used to recursively determine the
scalar integrals for $n>4$ as a linear combination of box integrals,
up to ${\cal O}(\e)$ corrections.
We also presented an approach to computing Feynman-parametrized
tensor integrals via the differentiation of the scalar integral
with respect to suitable combinations of the kinematic variables.
This approach also leads to simple differential equations for scalar
integrals, particularly box integrals.
In reference~[\OurPent] these general results are applied to the
specific computation of box integrals with massless internal lines,
but an arbitrary number of external masses, and to the computation
of pentagon integrals with all lines massless and with one
external mass.  The latter integrals are of use in the calculation
of one-loop contributions to amplitudes such as
$gg\to ggg$ and $Z\to q\bar q gg$.

We thank R. K. Ellis for discussions, especially regarding the
cancellation of the six-dimensional pentagon integral from tensor
integrals; and J. Vermaseren for helpful comments.

\vfill\eject\immediate\closeout\rfile
\centerline{{\bf References}}\bigskip\frenchspacing%
\input refs.tmp\vfill\eject\nonfrenchspacing

\bye